\documentclass[showpacs,apl,twocolumn,floatfix]{revtex4}

\usepackage{graphicx,float,amsmath}
\usepackage{mathrsfs}
\usepackage{natbib}
\pdfoutput=1
\begin{document}
\date{\today}

\title{Charge and spin transport  over record distances in  GaAs metallic n-type nanowires : I  photocarrier transport in a dense Fermi sea}

\author{H. Hijazi$^1$}
\author{D.~Paget$^2$}
\author{G. Monier$^1$}
\author{G. Gr\'egoire$^1$}
\author{J. Leymarie$^1$}
\author{E. Gil$^1$}
\author{F. Cadiz$^2$}
\author{C. Robert-Goumet$^1$}
\author{Y. Andr\'e$^1$}

\affiliation{%
$^1$ Université Clermont Auvergne, CNRS, SIGMA Clermont, Institut Pascal, F-63000 Clermont-Ferrand, France}

\affiliation{ 
$^2$ Physique de la mati\`ere condens\'ee, Ecole Polytechnique, CNRS, IP Paris, 91128 Palaiseau, France}



\begin{abstract}
We have investigated charge and spin transport in n-type metallic GaAs nanowires ($\approx  10^{17}$ cm$^{-3}$ doping level), grown by hydride vapor phase epitaxy (HVPE) on Si substrates. This was done by exciting the nanowire by tightly-focussed circularly-polarized light and by monitoring the intensity and circular polarization spectrum as a function of distance from the excitation spot. The spin-polarized photoelectrons give rise to a well-defined feature in the nearbandgap spectrum, distinct from the main line due to recombination of the spin-unpolarized electrons of the Fermi sea with the same minority photoholes. At a distance of 2 $\mu$m, only the main line remains, implying that photoelectrons have reached a charge  thermodynamic equilibrium with the Fermi sea. However, although no line is present in the intensity spectrum at the corresponding energy, the circular polarization is  still observed at the same energy in the spectrum, implying that photoelectrons have preserved their spin orientation and  that the two spin reservoirs remain distinct. Investigations as a function of distance to the excitation spot show that, depending on excitation power, a photoelectron spin polarization of $20 \%$ can be transported over a record distance of more than  20 $\mu$m.  This finding has potential applications for long distance spin transport in n-type doped nanowires.       
\end{abstract}
\maketitle

\section{Introduction}

The investigation of transport in systems of reduced dimensionality such as nanowires (NWs) is of interest both for fundamental reasons and for applications to solar cells \cite{krogstrup2013}, lasers  \cite{duan2003}, quantum computing  \cite{vandenberg2013} and spintronics \cite{krogstrup2013, duan2003, vandenberg2013}.  In silicon, time-resolved experiments have shown that photocarriers can be transported over $\approx 1$ $\mu$m \cite{gabriel2013}. For GaAs, the largest charge diffusion length is  of 4 $\mu$m at LT in quantum NWs \cite{nagamune1995}. However, most reported values at LT \cite{bolinsson2011, darbandi2016, gustafsson2010, spirkoska2009}  and RT \cite{gutsche2012} are in the submicron range. Finally, spin transport has to our knowledge been little investigated.\

N-type GaAs NWs on the metallic side of the Mott transition appear as a promising system for spin transport because of the large spin lifetime \cite{dzhioev2002, romer2010}. The efficiency of the Dyakonov-Perel process, which has been shown to be dominant at this doping level, is likely to be further reduced if the axial NW direction is $<$111$>$,  since the latter process is inefficient if the $k$ vector lies along $<$111$>$ \cite{Dyakonov1971}.\

At this doping level,  there appear tails in the valence and conduction band, due to statistical fluctuations of donor concentration \cite{efros1972, shklovskii1984, borghs1989}. It has been predicted that disorder in one dimensional semiconducting systems can lead to freezing of the spin relaxation \cite{echeverria2012}. On the other hand, it may be believed that the resulting potential fluctuations induce a localization of minority carriers and therefore strongly reduce the distance over which  minority carriers can be transported.\

In the present work and in its companion paper, hereafter refered to as [II], we use  metallic NWs of $\approx  10^{17}$ cm$^{-3}$ doping level, of exceptional quality and length, produced using Hydride Vapor Phase Epitaxy \cite{ramdani2010, gil2014, hijazi2019}. IN order to  investigate charge and spin transport along the NW, this NW is excited by a tightly-focussed circularly-polarized laser and  the evolution of the luminescence emission  spectrum and its polarization are monitored at 6K with spatial resolution along the NW. This allows us to investigate charge and spin transport along the NW. As shown before \cite{cadiz2014}, this approach has similarities with  time-resolved luminescence investigations \cite{redfield1970}, but is more appropriate to describe charge and spin transport. \ 

Metallic GaAs NW are potentially ideal candidates for charge and spin transport because of  two distinct mechanisms, which have been little investigated in the past. Firstly,  it will be shown in [II] that charge can be transported in the bandtails over lengths as  large as  20 $\mu$m because of the presence of large internal electric fields of ambipolar  origin. Secondly, in the present paper, we investigate, as performed before for bulk materials \cite{knox1988} and heterostructures \cite{brill1994}, to what level the presence of a Fermi sea of spin-unpolarized intrinsic electrons affects charge and spin transport in the NW. It is shown that thermodynamic equilibrium between photoelectrons and the Fermi sea is established after a small distance of 2 $\mu$m, but that the two spin reservoirs remain distinct up to 20 $\mu$m.\

This conclusion is reached using a spatially-resolved investigation of the luminescence and polarization spectra. Besides the main emission, due to recombination of photoholes with the Fermi sea, a narrow polarized feature is found corresponding to recombination of spin-polarized photoelectrons lying near the Fermi level with the same photoholes. This line disappears within a distance of 2 $\mu$m from the excitation spot, thus revealing the establishment of equilibrium between the charge reservoirs of photoelectrons and intrinsic electrons. At the same energy in the spectrum, the circular polarization is still present  up to the maximum distance of 2 $\mu$m, revealing that spin equilibrium has not been reached. Moreover, spatially-resolved investigation of the luminescence degree of circular polarization shows that spin orientation  can be transported up to the maximum distance of  20 $\mu$m, thus revealing spin transport over record lengths. At this distance, depending on excitation power, the photoelectrons can have a spin polarization as large as $20 \%$. This shows the potential interest of the present NW for spin transport.\

This paper is organized as follows. The following section is dedicated to a theoretical background and to the experimental details. In Sec. III, the spatially-resolved luminescence and polarization spectra are presented. These results are interpreted in Sec. IV.\

\section{Principles}
\label{exp}

\subsection{NW growth and preparation}

Here, we study  gold-catalyzed NWs, HVPE-grown on Si(111) substrates at 715  $^\circ$C. These NWs have a length of several tens of $\mu$m and are characterized by a  pure zinc blende structure,  free of polytypism and cristalline defects \cite{ramdani2010, gil2014}. Since the HCl flux injected inside the reactor produces  SiCl$_4$ which acts as a doping precursor, the NW have a donor doping level $N_D$ in the  low  $ 10^{17}$ cm$^{-3}$ range, weakly dependent on NW diameter \cite{hijazi2019}. This value is about one order of magnitude larger than the one of the Mott transition \cite{benzaquen1987, cutler1969, anderson1958}. This value has been obtained from an analysis of the shape of the luminescence spectrum and by Raman analysis \cite{goktas2018} and has been confirmed using a mapping of the luminescence intensity, leading to the conclusion that the surface depletion layer has a thickness of the order of 90 nm (see supplementary material).\ 

Immediately after growth, the NW were introduced  without air exposure into a UHV chamber  and were treated at  300K by a nitrogen plasma produced by a commercial electron cyclotron resonance source (SPECS MPS-ECR) operating in atom mode at a pressure of $2.5.10^{-5} $ mbar and described elsewhere \cite{mehdi2018}. In order to obtain a homogeneous nitridation on the  NW surface, the angle between the source and the substrate surface was kept  at 45$^\circ$ for 1h and at -45$^\circ$ for 1h. This method enables to produce a thin layer of nitride at the GaAs surface which reduces the surface oxidation under air exposure and the surface recombination velocity \cite{mehdi2019, andre2020}.\

The NWs, standing on the substrate, were mechanically deposited horizontally on a grid of lattice spacing 15 $\mu$m. An optical microscope was used to note the coordinates  of the individual  NW. As found by scanning electron microscopy, the NW used here had a  length of $  80$ $\mu$m and a diameter of $\approx 220$ nm.\

\subsection{Background on luminescence of metallic n-type GaAs}

\begin{figure}[tbp] \centering
\includegraphics[clip,width=8.6 cm] {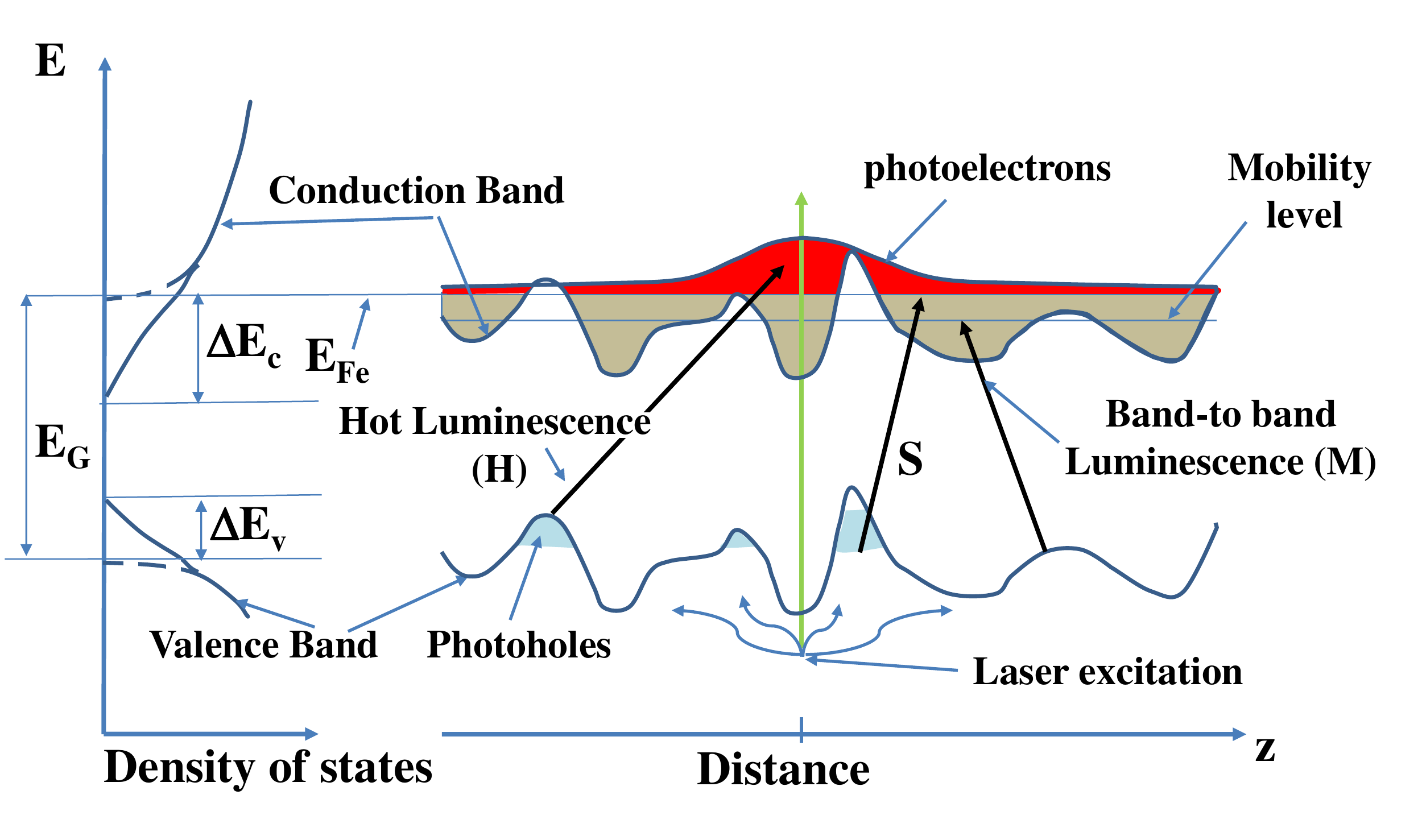}
\caption{Scheme for carrier excitation in the potential fluctuations of the conduction and valence bands of n-doped NW. Intrinsic electrons occupy the fluctuations of the conduction band up to the quasi Fermi level $E_{Fe}$. With the value of the doping level, this Fermi level lies above the mobility level, above which the electrons are no longer confined and participate to the electric conductivity. The various nearbandgap emissions are labelled in the same way as in the spectra of Fig.\ref{Fig4} (see text).}
\label{Fig1}
\end{figure}

Shown in the right panel of Fig. \ref{Fig1} are  the spatial potential fluctuations of the bottom of the conduction band and of the top of the valence band for n-type metallic GaAs. These fluctuations originate from statistical fluctuations of the donor concentration. In  a sphere of radius $R$, the statistical fluctuation of the mean number of donors N, given by $N= 4\pi N_D/(3R^3)$,  is $\sqrt{N}$, so that the potential fluctuation  is $\sqrt{N} q^2/(\epsilon \epsilon_0 R)$ where $\epsilon$ is the static dielectric constant, $\epsilon_0$ is the permittivity of vacuum  and $q$ is the absolute value of the electronic charge \cite{efros1972}. The potential fluctuations are screened by mobile carriers. Within the Thomas Fermi (TF) 3D model the screening concerns fluctuations of extension  larger than $r_s =(1/2) \sqrt {a_0^* N_D^{1/3}} \approx a_0^* $ where $a_0^*$ is the donor Bohr radius \cite{desheng1982, lanyon1979, note10}. 

Calculations of the energy dependence of the conduction and valence band densities of states have been performed in the past \cite{lowney1986}.  For $ N_D \approx 10^{15}$ cm$^{-3}$ range,  one merely observes a broadening of the donor band. In contrast, for the present doping level $ N_D \approx 10^{17}$ cm$^{-3}$,  the fluctuations generate a tail lying lower than the conduction band ( see left panel of  Fig. \ref{Fig1}). The amplitude of this tail has been found  of the order of $\Delta E_c \approx 8$ $meV$,  i. e. comparable with the donor binding energy, while the  density of states $\rho_c(\epsilon_c)$ increases linearly  as a function of energy $ \epsilon_c$ with respect to the bottom of the tail \cite{lowney1986}. Conversely, the valence band also exhibits a  tail of amplitude $\Delta E_v \approx 8$ meV, with a density of states $\rho_v(\epsilon_v)$ increasing  also linearly with increasing energy   with respect to the top of the tail $ \epsilon_v$.\

It has been found that the dynamic properties of the two types of carriers in the fluctuations are very different \cite{levanyuk1981, arnaudov1977}. Under light excitation,  the electronic reservoir is characterized by a thermodynamic equilibrium defined  by  a quasi Fermi level since, because of the small electronic mass, electronic diffusion by tunnel processes from one well to the other one is quite efficient. On the other hand, photoholes tend to get trapped in the  potential wells, since  relaxation of their kinetic energy  occurs in a short characteristic time of 1 ps  \cite{chebira92, bimberg1985}, where tunneling processes are less probable because of their large effective mass. As a result, as verified experimentally \cite{redfield1970}, the holes cannot be described by a thermal equilibrium, but by a balance between thermalization and recombination. The valence band levels are occupied by holes with an occupation probability given by  

\begin{equation} \label{holedist}
	f_v(\epsilon_v)  =  \frac{W_p p}{W_p p +W_n (n_0+n)} F(E_{Fh}).	
	\end{equation} 
	
Here $W_p$ is the capture probability of a hole, $W_n$ is the probability for recombination with electrons and  $ F(E_{Fh})$ is a Fermi function of $\epsilon_v$ with an effective Fermi energy $E_{Fh}$  given by   

\begin{equation} \label{holeferm}
	E_{Fh}=k_B T_h ln \left[ \frac {W_p N_v}{W_p p +W_n (n_0+n)}\right] \approx k_B T_h ln \left[ \frac {W_p N_v}{ W_n n_0}\right]
	\end{equation}		
\noindent
Here $n$ and $n_0$ are the concentrations of photoelectrons and intrinsic electrons, $p$ is the hole concentration and  $N_v$ is the valence band effective density of states. The approximate expression is valid at low excitation power, for which $n<<n_0$ and $W_p p << W_n n_0$. The hole distribution differs from a  Fermi one because of the concentration dependence of the prefactor in Eq. \ref{holedist}. It has been proposed that, because of the large recombination probability of holes at the top of the fluctuations, the hole energy distribution is narrow  and peaks at some intermediate kinetic energy \cite{desheng1982, arnaudov1977}.\

The  luminescence intensity  at energy $E$ of the main line, due to recombination between photoholes and intrinsic electrons, is  proportional to  

\begin{equation} \label{IMind}
	I_{main} (E) =  \int_{0}^{\infty} W(\epsilon_c, \epsilon_v) \rho_c(\epsilon_c)  \rho_v(\epsilon_v) f_c(\epsilon_c ) f_v(\epsilon_v ) d \epsilon_c.	
	\end{equation} 
\noindent	
with  $E  = E_G  - \epsilon_c - \epsilon_c$ and where k-conservation does not occur because of disorder \cite{desheng1982}. Here  $f_c$ is the electron  occupation probability and $ W(\epsilon_c, \epsilon_v)$ is the recombination probability. Expressions of these quantities have been given in Ref \cite{levanyuk1981}, which  reports a calculation of the shape of the luminescence spectrum. \ 

At a given radial position $r$ and axial position $z$ in the NW, the intensity, obtained by integration of Eq. \ref{IMind} over energy $E$, is of the form   
\begin{equation}\label{Imain}
	 I_{main}= K n_0p. 
	\end{equation} 	
where $K $ is the bimolecular recombination constant. In the same way, the intensity of the emission due to recombination between photoelectrons and photoholes is  
\begin{equation}\label{Ihot}
	 I_{hot}= K_{hot} np. 
	\end{equation}
\noindent
where $K_{hot}$ is the corresponding bimolecular recombination constant. The detected luminescence intensity at a given axial position $z$ along the NW is obtained by averaging over the radial coordinate $r$.

The luminescence spectrum may also contain features due to distinct recombination processes. Several works have reported a composite structure of the nearbandgap emission, attributed to a residual excitonic signal  \cite{desheng1982} or to band-to-band recombination, while the main line is attributed to donor-related recombination \cite{lee1996}. Independently, it has been shown that, because of screening of the electron-hole interaction, excitons cannot exist for the present doping level, since the exciton absorption peak disappears for  $ N_D > 10^{16}$ $cm^{-3}$ \cite{shah1977}. However, it has been shown that biexcitons can survive the Mott transition \cite{Shahmohammadi2014}.
 
\subsection{Experimental}
  
\begin{figure}[tbp]
\includegraphics[clip,width=8.6 cm] {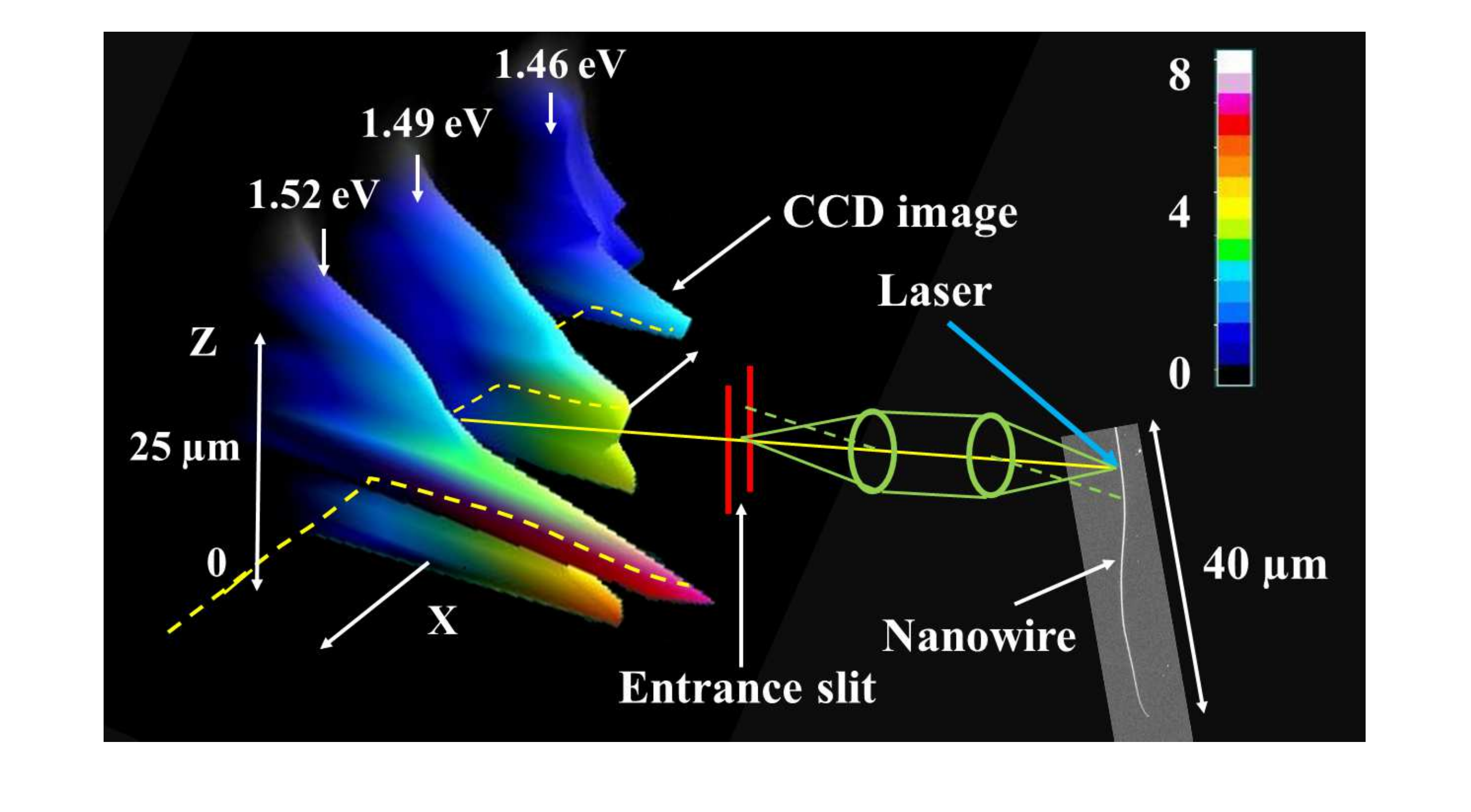}
\caption{Scheme of the experimental setup, showing a scanning electron microscope picture of the NW and a  3D picture of the CCD image at 6K for an excitation power of $9$ $\mu$W. The image exhibits 3 main emissions, including the nearbandgap luminescence near 1.52 eV and two less intense bands due to recombination at residual acceptors. Section of this image along the X axis, perpendicular to the entrance slit (dotted curve), gives the luminescence spectrum at a given position in the NW.}
\label{Fig2}
\end{figure}

The experimental setup is sketched in Fig. \ref{Fig2}. The excitation light is a tightly-focused, continuous-wave, laser beam (Gaussian radius $\sigma \approx 0.6\; \mu$m, energy $1.59$ eV).  The luminescence light is focused on the entrance slit of a spectrometer equipped with a CCD camera as a detector. For a n-type material, the band-to-band emission is dominated by recombination of minority photoholes with intrinsic electrons.\
For spatially-resolved spectral analysis, one monitors the image from the CCD detector. A typical image, taken for a NW temperature of 6K, is shown in Fig. \ref{Fig2} for an excitation power of 9 $\mu$W  \cite{note8}.  Here,  the NW  is adjusted so that its image by the detection optics is parallel to the spectrometer entrance slit (axis Z). Thus, section of the image along the perpendicular axis X gives the luminescence spectrum at the corresponding position on the NW. As shown in Fig. \ref{Fig2},  the spatial profiles extend well beyond the zone of optical excitation ($\approx 0.6$ $\mu$m), so that the monitoring of the spectra as a function of distance gives information on evolution of the photocarrier charge and spin reservoirs during transport away from the excitation spot.\

Liquid crystal modulators were used to circularly-polarize the excitation laser  ($\sigma^{\pm}$-helicity), in order to generate spin-polarized photoelectrons and to selectively detect the intensity $I(\sigma ^{\pm})$ of the luminescence components with $ \sigma ^{\pm}$ helicity. Since photoholes as well as intrinsic electrons  are spin-unpolarized, the band-to-band luminescence is expected to be also circularly-unpolarized. Conversely, for recombination with spin-polarized photoelectrons,  one monitors the difference signal 
\begin{equation} \label{Idif}
	I_{D}= I_{hot}(\sigma ^+) - I_{hot}(\sigma ^-)  = K_{hot} \mathscr{P}_i s.	
	\end{equation} 
\noindent	
where $ \mathscr{P}_i= \mp 0.5$ for $ \sigma ^{\pm}$- polarized excitation. This signal is related to the photoelectron spin density  $s=n_+ - n_-$, where $n_{\pm} $  are the concentrations of photoelectrons with spin $\pm 1/2$, choosing the direction of light excitation  as the quantization axis. Finally, the ratio $\mathscr{P}= I_{D}/I_{S}$ is defined as the degree of circular polarization of the luminescence and is  $\mathscr{P}=  \mathscr{P}_i s/n$.\

\section{Spatially-resolved sum and difference  spectra}

\subsection{Spectral analysis at the excitation spot}

\begin{figure}[tbp]
\includegraphics[clip,width=8.6 cm] {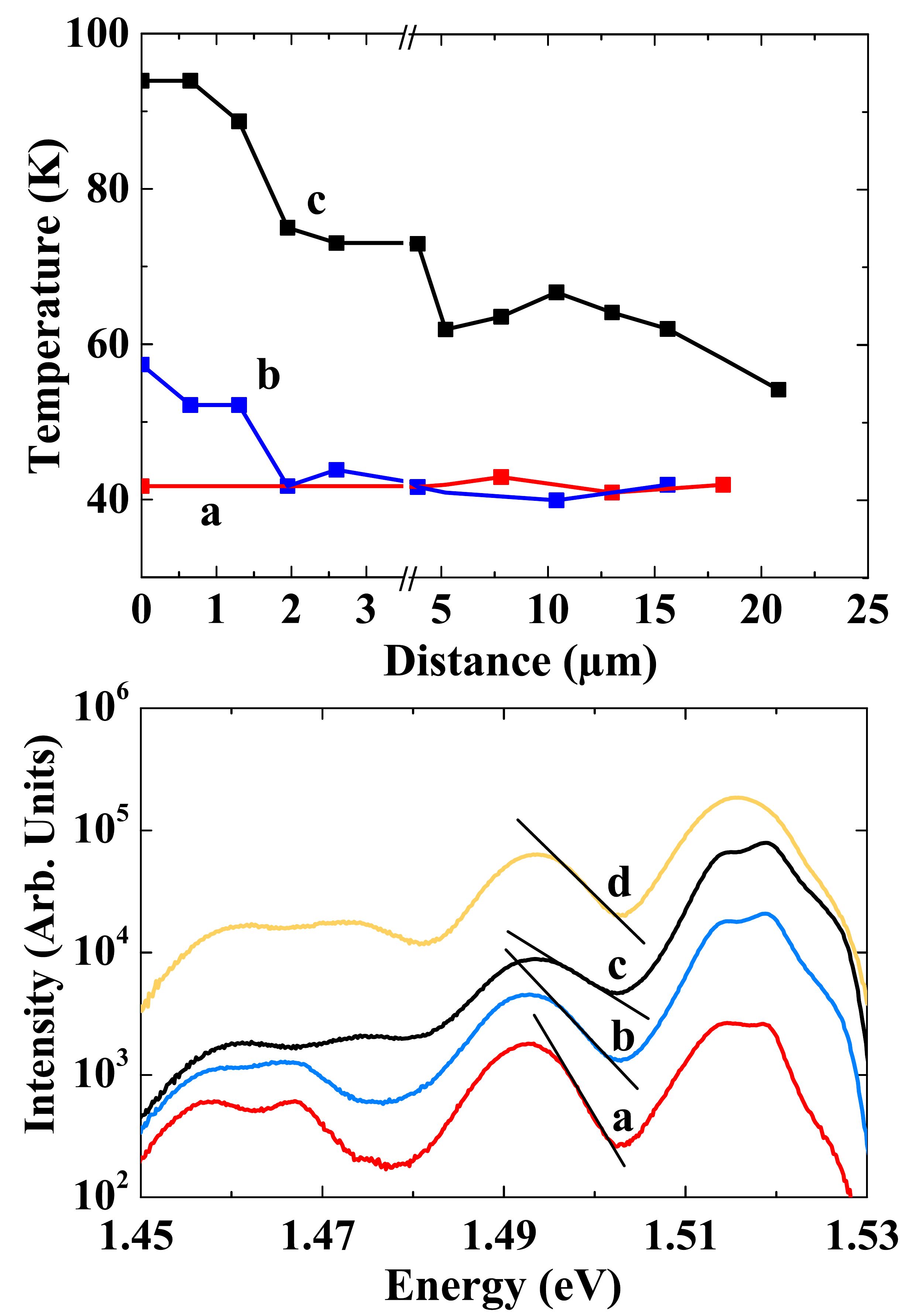}
\caption{The bottom panel shows in logarithmic units the intensity spectrum, at the place of excitation and  for an excitation power of 9 $\mu$W (a), 45 $\mu$W (b) 180 $\mu$W (c). Also shown in Curve d is the spectrum for an excitation power of 180 $\mu$W (same as Curve c), but at $z=3.12$ $\mu$m.  From the slope of the high-energy side of the acceptor-related luminescence  at 1.493 eV, one obtains the temperature of the electron gas, $T_e$.  The top panel shows, for the same excitation powers, the dependence of $T_e$ as a function of $z$. Note the change of linear scale of the X axis at  $3$ $\mu$m.}
\label{Fig3}
\end{figure}

  The luminescence image shown in Fig. \ref{Fig2} consists in three bands. Besides the   nearbandgap luminescence near 1.52 eV,  the  band related  at 1.49 eV is due to residual carbon acceptors \cite{kisker1983, skromme1984}. The band near 1.46 eV, possibly caused by carbon acceptors perturbed by nitrogen atoms originating from the surface passivation \cite{leymarie1989,ruhstorfer2020} has properties very close to the former one. The bottom panel of Fig. \ref{Fig3} shows, in logarithmic units, the luminescence intensity spectra at $z=0$ for increasing excitation powers (Curves a, b and c, corresponding to 9 $\mu$W, 45 $\mu$W and 180 $\mu$W  respectively) and the spectrum for an excitation power of 180 $\mu$W but for  $z=3.12$ $\mu$m (Curve d).\

We first consider the acceptor emission at 1.492 eV. The lineshape does not seem to be affected by disorder. As realized earlier \cite{desheng1982, levanyuk1981}, this is because of the relatively small value of the acceptor Bohr radius, so that disorder causes a negligible broadening. It has also been shown that the acceptor luminescence originates from recombination of electrons at the Fermi level, so that the shift of the acceptor emission because of disorder,  of   $E_{Fe} -E_c$  \cite{levanyuk1981}, is very small \cite{note9}. The high-energy slope of the  corresponding peak  increases with excitation power (Curves a-c of Fig. \ref{Fig3}) and becomes smaller at  $z=3.12$ $\mu$m (Curve d). As already recognized before \cite{ulbrich1989, weisbuch1977}, this slope is related to the electron temperature $T_e$ which is, in the present case, the temperature of the Fermi sea.  A fit of this line with a shape of the form exp$(-h\nu/k_BT_e)$ directly gives  $T_e$. The top panel shows the  corresponding spatial profiles of $T_e$. At very low power, $T_e \approx 40K$ and  is, as expected, independent on distance. The temperature at $z=0$ increases with excitation power, as seen from a comparison between Curves a, b, and c.  Fora power of 180 $\mu$W, which is  the highest power at which the line can be resolved from the nearbandgap emission (Curve c), $T_e$ can be as large as 95K at the place of excitation. In this case, as seen in Curve d, $T_e$ decreases with distance to the excitation spot. \ 

\begin{figure}[tbp]
\includegraphics[clip,width=8.6 cm] {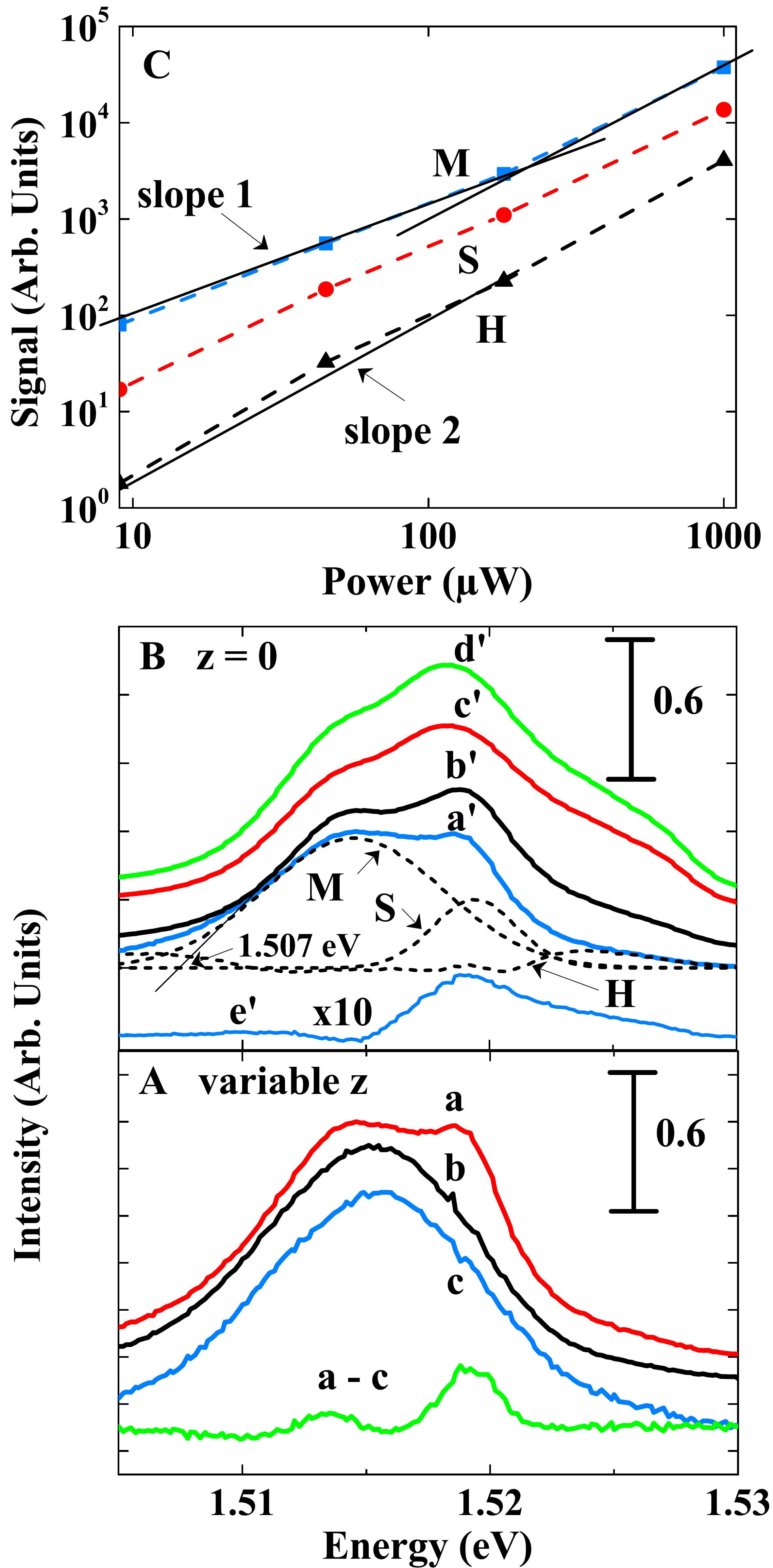}
\caption{Panel A shows the intensity spectra of the nearbandgap line for an excitation power of 9 $\mu$W (Curve a), of 45 $\mu$W (Curve b), and  180 $\mu$W (Curve c). The distance $z$ was  changed so that  all these spectra have identical maximum signals. Its value was $z=0$  for Curve a, $z=5.2$ $\mu$m  for Curve b and $z=15.6$ $\mu$m  for Curve c. Since line S at 1.519 eV is only visible in Curve a, that is near the excitation spot, this signal is not directly related to the photocarrier concentration, but rather to the distance from the excitation spot. Panel A also shows the difference a-c, which contains mainly  line S.  Panel B shows the corresponding spectra  at a fixed distance ($z=0$) of the nearbandgap line for an excitation power of 9 $\mu$W  (Curve a), of 45 $\mu$W (Curve b),  180 $\mu$W (Curve c) and 1 mW (Curve d). Curve e' shows the difference spectrum (x10), given by Eq. \ref{Idif} and related to the spin orientation, at an excitation power of 9 $\mu$W. The circular polarization  on this spectrum is mostly limited to lines S and H, with a weak polarization on line M. Panel C shows, in logarithmic units, the power dependences of the  intensities of lines M, S and  H, as obtained from a decomposition of the spectra of Panel B.}
\label{Fig4}
\end{figure}

The nearbandgap  normalized luminescence intensity spectra at $z=0$ are shown in  Panel B of Fig. \ref{Fig4} for selected excitation powers. These spectra are composed of a main line near 1.515 eV labelled M, of a shoulder at 1.519 eV labelled S and of a high-energy tail, above 1.52 eV, labelled H. Curve e' of Panel B  shows the difference spectrum for the smallest excitation power of 9 $\mu$W. One sees that the polarization of line M is small, so that this line is due to recombination of  spin-unpolarized photoholes with  intrinsic electrons. Indeed, because of band filling, electrons  lying below the Fermi level cannot be spin-polarized. Conversely, lines S and H are polarized and are therefore due to recombination of spin-polarized photoelectrons at the photoelectron quasi-Fermi level and above this level, respectively.\

For quantitative analysis, the  spectra were decomposed into elementary contributions, as shown for Curve a' of Panel B of Fig. \ref{Fig4}.  Line S was fitted by a  gaussian component of half-width 2.3 meV and peak energy 1.5195 eV. The  width of  line S is  relatively small since this line reflects the joint widths of the photoelectron distribution, determined by the temperature and of the photohole distributions which, as shown in Sec. IIB, is relatively narrow. The position  of line S, which corresponds to the difference between electron quasi-Fermi level and the hole energy, is found to depend very weakly on  excitation power, in agreement with the expression of  $E_{Fh}$, given by Eq. \ref{holeferm} at weak excitation power.  For  line M, one has used  a gaussian shape of half width $\approx 6$ meV, i. e. comparable with values measured elsewhere on Si-doped NWs  \cite{ruhstorfer2020}. Line M is broader than line S, since its width is determined by the width of the Fermi sea, of the order of the electron Fermi energy. Line M is extrapolated at low energy to a value of  1.507 eV, in relatively good agreement with the value of 1.503 eV expected from Ref.  \cite{lowney1986} for this doping level. Finally, the hot photoelectron contribution H was taken as the  residual signal, obtained by subtracting components S and M from the experimental profile. The shape of this component was found to depend weakly on excitation power.\

Shown in Panel C of  Fig. \ref{Fig4} are the power dependences of the integrated intensities of  lines M, S and H. As expected, the intensity of line M is proportional to the excitation power, because of the linear dependence of $p$ on excitation power in Eq. \eqref{Imain} (monomolecular recombination). Conversely, that of line H is proportional to its square, since in Eq. \eqref{Ihot}, both $p$ and $n$ increase with excitation power (bimolecular recombination). Note that the exponent of the increase of the intensity of line S, of 1.4, is slightly smaller than the value of 2 expected from Eq. \eqref{Ihot}. This departure may be due to a power dependence of $K_{hot}$ or to the fact that  a power-dependent fraction of the photoelectrons is already incorporated into the Fermi sea at $z=0$.\ 
   
\begin{figure}[tbp]
\includegraphics[clip,width=8.6 cm] {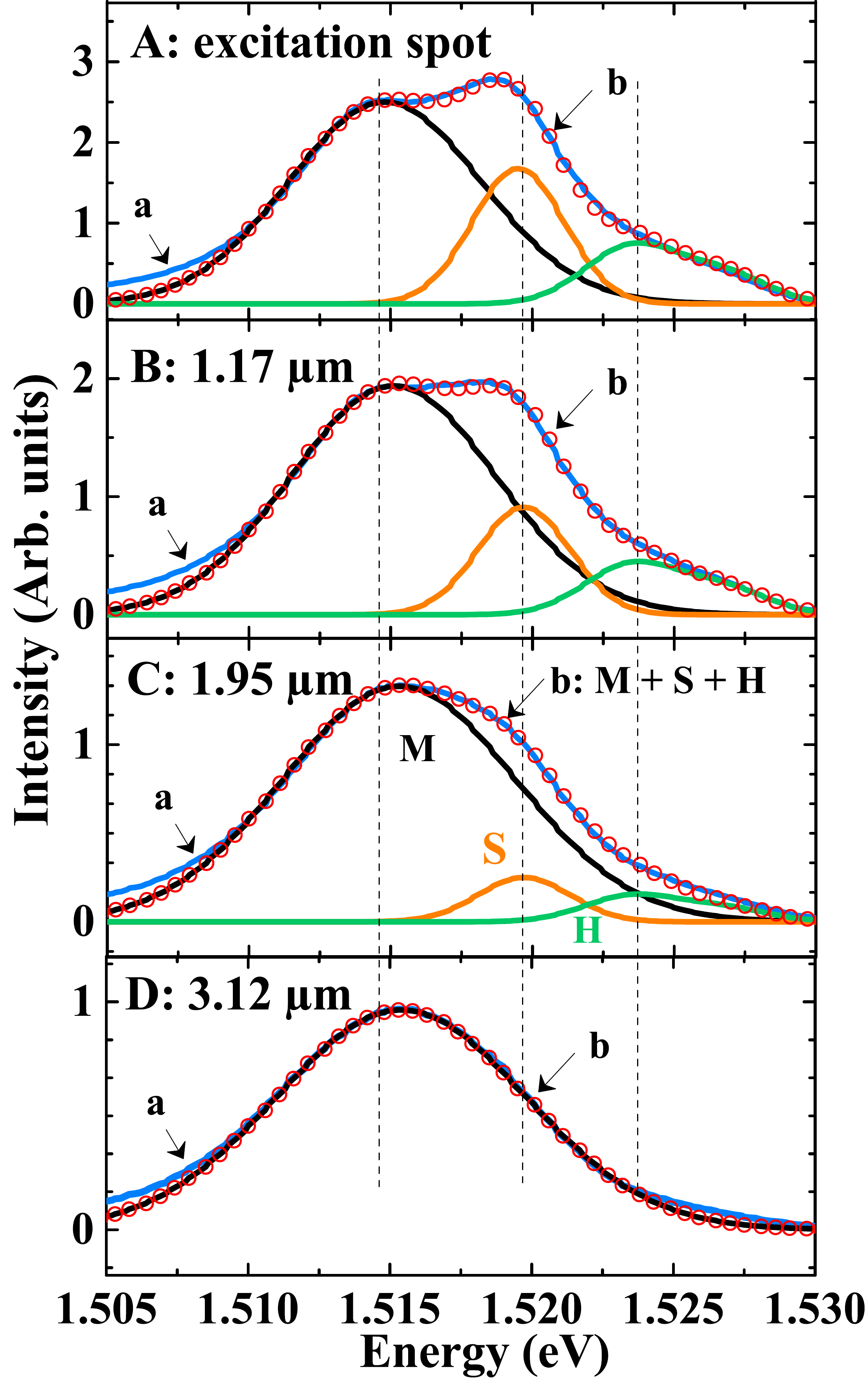}
\caption{Spatially-resolved  spectra (Curves a) at the excitation spot (A) and for selected distances from this spot of 1.17 $\mu$m (B), 1.95 $\mu$m (C) and 3.12 $\mu$m (D), for an excitation power of $45$ $\mu$W. All the spectra were decomposed  using the main component (M), the shoulder at 1.519 eV (S) and the hot electron  contribution (H). Curves b show the sum of these contributions and closely follow the experimental spectra. }
\label{Fig5}
\end{figure}

\subsection{Intensity spectra  for selected distances to the excitation spot }
\begin{figure}[tbp]
\includegraphics[clip,width=8.6 cm] {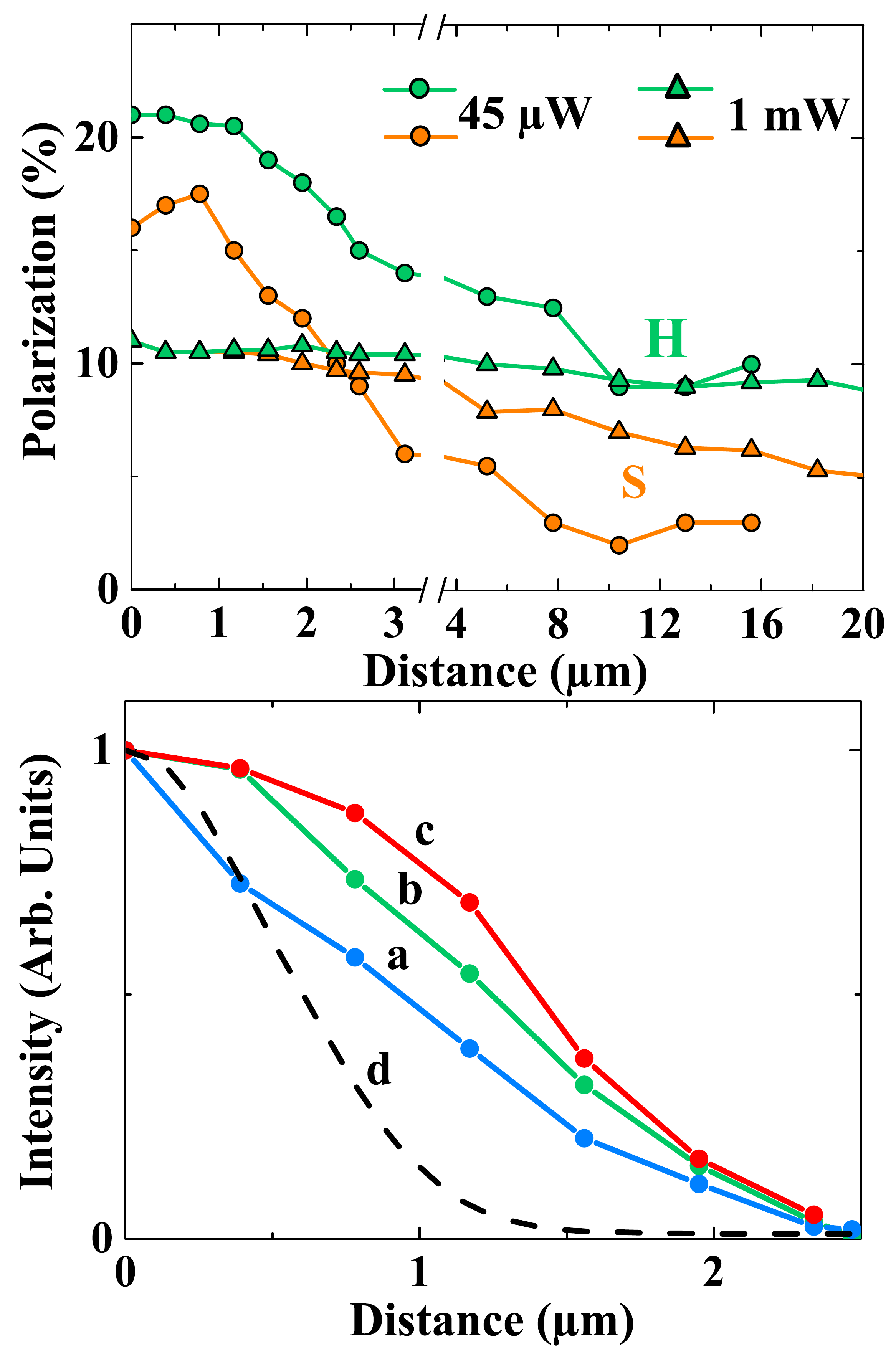}
\caption{The bottom panel shows the spatial profiles of the  intensity of line S in the intensity spectra, obtained using the decomposition shown in Fig. \ref{Fig4}, for an excitation power of 9 $\mu$W (Curve a),  45 $\mu$W (Curve b) and 1 mW (Curve c). This decay reveals the establishment of thermodynamic equilibrium between photoelectrons and intrinsic electrons. Also shown in Curve d is the  laser intensity spatial profile. The top panel shows the spatial profiles of the degree of circular polarization  for line H  and line S for an excitation power of  45 $\mu$W (open symbols) and 1 mW (closed symbols).}
\label{Fig6}
\end{figure}

Fig. \ref{Fig5} shows the intensity  spectra for an excitation power of   $45$ $\mu$W and for selected distances to the excitation spot. These spectra were decomposed into the elementary contributions of lines M, S and H, in the same way as for Curve a' of Fig. \ref{Fig4}, and the spatial profiles of line S are shown in  Fig. \ref{Fig6} for several excitation powers. Fig. \ref{Fig5} shows that line S  disappears over a characteristic distance of $ \approx 2$ $\mu$m so that the spectrum shown in Panel D mostly exhibits  line M, with a weak residual H signal above 1.52 eV. The decay is slower than that of the laser spatial profile, shown in Curve d of Fig. \ref{Fig6}, implying that the changes in these spectra are not directly related to the photocarrier creation rate but to evolution of the photocarrier system during transport. \

Quite generally, the disappearance of line S can be attributed to a nonlinear effect caused by the decrease of photocarrier concentration (note the decrease of the emission intensity by a factor of 3  between Panels A and D of Fig. \ref{Fig5}), or to the effect of transport on the photocarrier system. In order to test the sole effect of distance to the excitation spot, we compare in the bottom panel of Fig. \ref{Fig4}  the intensity spectra as a function of distance, where  for each distance the excitation power  was adjusted in such a way that the emission intensity and therefore the carrier concentrations are constant. Curve a is identical to Curve a' of Panel B, taken at $z=0$ and for the smallest excitation power of 9 $\mu$W. Curves b' and c' are taken for  distances of $z=5.2$ $\mu$m and $z=15.6$ $\mu$m but for excitation powers of 45 $\mu$W and 180 $\mu$W, respectively. \ 

As seen from the difference between the spectrum at the excitation spot (Curve a) and the spectrum taken at $15.6$ $\mu$m (Curve c), the intensities of lines H and M are quite similar between Curves a, b, and c, implying that they are related to the photocarrier concentrations rather than to the distance from the excitation spot. In contrast, line S is absent from  Curve b and Curve c of Fig. \ref{Fig4} so that its relative magnitude depends on the distance to the excitation spot rather than on the photocarrier concentration. This allows us to exclude, at variance with earlier work \cite{desheng1982, lee1996, Shahmohammadi2014}, electronic species such as excitons, biexcitons  as the origin of line S, since in this case the magnitude of this line should mostly depend on carrier concentration and therefore on the intensity of line M. These results rather show the relevance of irreversible establishment of equilibrium occuring after generation of electron-hole pairs during transport away from the excitation spot. This equilibrium concerns the photoelectron gas since since the hole energy relaxation time is quite short and smaller than 1ps  \cite{chebira92} and since establishment of equilibrium among the hole gas would  also affect line M. \

\subsection{Difference and polarization  spectra for increasing distances to the excitation spot}

\begin{figure}[tbp]
\includegraphics[clip,width=8.6 cm] {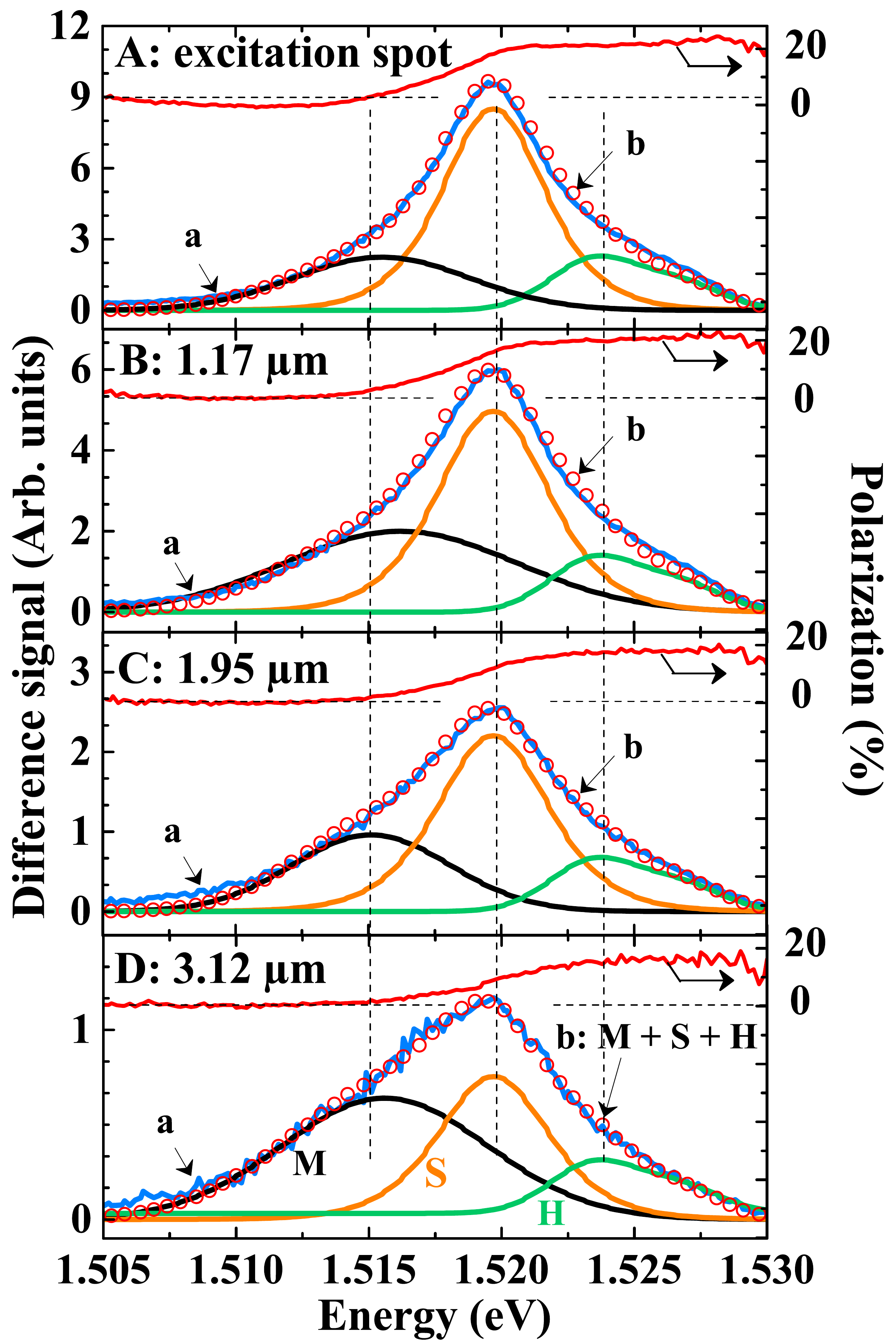}
\caption{Same as Fig. \ref{Fig5}, but for the difference spectra. Also shown are the polarization spectra, defined as the ratio between difference signal and intensity. Note that component S which has disapeared at large distance from the intensity spectrum (Panel D), is still the dominant feature of the corresponding difference spectrum.}
\label{Fig7}
\end{figure}

Fig. \ref{Fig7} shows the difference spectra in the same conditions as Fig. \ref{Fig5} as well as, for each distance, the polarization spectra.  It is seen that, at large distance, there persists a significant S line in the difference spectrum although no specific feature is detected in the corresponding intensity spectrum. This finding implies that, in spite of the establishment of a charge equilibrium, the photoelectrons and intrinsic electrons still form two distinct spin reservoirs. The persistence of a significant Fermi edge spin polarization  at  large distance reveals the weakness of the spin relaxation processes \cite{dzhioev2002}.\

It is then assumed that each photoelectron  spin reservoir, of spin $\pm$, has reached an internal equilibrium characterized by a  Fermi energy $E_{F\pm}$,  such that $E_{F+} +  E_{F-} =2E_{F}$, in order to ensure charge equilibrium. Developing the function  $s \approx F(E_{F+})- F(E_{F-})$ to first order in $E_{F+}-E_{F-}$ and dividing by the electron concentration, one finds that the polarization is proportional to  $1-F(E_{F})$. The polarization spectrum in Panel D can be perfectly approximated by $1-F (E_{F})$, from which we find that the difference between the electron quasi-Fermi energy $E_{F}$ and the hole energy is of 1.5195 eV. The fact that this energy coincides with the energy of line S is a further confirmation that this line is due to recombination of photoelectrons at the quasi Fermi level.\ 

For analysis of the polarization spatial profiles of lines S and H beyond  $z=3$ $\mu$m, it was chosen to monitor  the values in the polarization spectra at the respective energies of the peaks of the two lines. This procedure enables to determine the profiles  independently of decomposition of the sum and difference spectra up to the maximum distance of  $z= 20$ $\mu$m even if no line S is present in the intensity spectrum. The resulting polarization spatial profiles  are shown in the top panel of Fig. \ref{Fig6}. At the excitation spot and for the smallest excitation power, the luminescence polarization for hot electrons and to some extent for Fermi edge electrons  is close to the maximum value of $ 25 \%$ without losses by spin relaxation. For the maximum excitation power,  the polarization of line H is $10\%$. This polarization keeps the same value independently on distance, while the polarization of Fermi edge electrons slowly decreases with distance and is  $5\%$ for $z= 20$ $\mu$m. For a reduced  excitation power of 45 $\mu$W,  the polarization of line H  of $20  \%$, decreases to $ 10  \%$, these values being  $16 \%$ and  $ 3 \%$ for line S.  For the smaller excitation power, in agreement with Fig. \ref{Fig6}, the polarization of line S stays larger than $ 10 \%$  up to  $z= 3 $ $\mu$m, and subsequently decreases to $\approx 3 \%$ for $z= 20$ $\mu$m.\

In summary, it has been found that spin transport can occur over record lengths,  with a luminescence  polarization of  $ \approx 10  \%$ at $z= 20$ $\mu$m for hot photoelectrons. For Fermi edge photoelectrons, spin transport occurs over similar distances, but the polarization losses are larger, in particular for a small excitation power. Even for a low excitation power, polarization of Fermi edge electrons is larger than $ 10  \%$ up to $z= 3$ $\mu$m, at which charge equilibrium  with the unpolarized Fermi sea is established.\

\section{Discussion }

In order to outline the physical process at the origin of the experimental effects, we first   estimate the order of magnitude of the time for transport out of the laser spot. Taking a typical value of the diffusion constant  of $100$  cm$^2/s$ \cite{cadiz2014}, one  finds that this time ,  $ \approx \sigma^2/D$ is of the order of several tens of ps. In the same way, with the excitation power used for Fig. \ref{Fig5}, we estimate that the photocarrier concentration at the excitation spot is of the order of $10^{14}$ cm $ ^{-3}$. Although this estimate is very approximate, it allows us to conclude that the photocarrier concentration is much smaller than the doping level.\

\subsection{Establishment of charge equilibrium}


The first dynamic process which occurs after creation of an electron in the conduction band  is emission of an optical phonon. This emission has been found to occur in a time of  $\approx 0.2$ ps i. e. significantly smaller than the time for establishment of the equilibrium  \cite{kash1985}. Although this time may be larger at high excitation power  because of screening of the electron-phonon interaction \cite{seymour1980}, the observation of a significant S signal at $z=0$ suggests that emission of optical phonons is complete before diffusion out of the excitation spot.\

Electron-electron collisions are also known to enable efficient establishment of equilibrium among the electron gas. The time for collisions between  electrons  has been calculated including screening  by an electron hole plasma and found smaller than 1ps independently on concentration and temperature \cite{binder1992}. Thus it may be believed that establishment of equilibrium among  photoelectrons occurs  before they leave the excitation spot.\ 

The  present experimental results show that, in contrast, equilibrium between the photoelectrons and Fermi edge intrinsic electrons only occurs after  a distance of $2$ $\mu$m. Experimentally, the time for establishment of equilibrium between photoelectrons and a Fermi sea of electrons  has been found to be shorter than 30 fs. However, this was found in a modulation-doped structure i.e. without screening by charged donors \cite{knox1988}. The slower establishment of equilibrium with the Fermi sea is consistent with the experimental finding that the interaction between photoelectrons and the Fermi sea, rather than  occuring through single particle processes, modifies the equilibrium of the overall Fermi sea \cite{brill1994}. This suggests that, at the excitation spot, the equilibrium of  the photoelectron reservoir may be a Boltzmann-like with a temperature distinct of that of the Fermi sea. Further evolution consists in equalization of the two temperatures. The relaxation of the photoelectron temperature towards equilibrium has been shown to be much slower than the above processes and to occur in a characteristic time larger than several tens of ps   \cite{delfatti1999} i. e. comparable with the above estimate for transport up to 2 $\mu$m.\

As seen from Fig. \ref{Fig6}, the characteristic distance for establishment of equilibrium relatively weakly depends on excitation power and increases by less than a factor of 2 between Curves a and c, while the excitation power has increased by two orders of magnitude. The fact that the resulting increase of the heat capacitance of the photoelectron reservoir  has little effect on the photoelectron dynamics suggests that, even for the maximum power,  the photoelectron concentration is  smaller than that of the Fermi sea. The slowing down of the interaction between the two types of reservoirs could be caused  by screening of the interactions between electrons by mobile charges \cite{cadiz2015b}.\

 
\subsection{Spin dynamics during transport}

At the excitation spot, the polarization for the highest excitation power is  smaller than for the lowest excitation power. We believe that these losses are due to exchange with photoholes (Bir Aronov Pikus mechanism  \cite{bir1975}). They should indeed be larger at high excitation power, at which the hole concentration at the excitation spot is significant.\

Away from the excitation spot, and except for hot electrons at the maximum excitation power, the polarization decreases with $z$, implying that  photoelectrons undergo some polarization losses. In order to explain these effects, it is recalled that in this range, the dominant relaxation mechanism is the D'yakonov Perel one. The relaxation time is usually given by $1/T_1= \Omega^2 \tau _c$, where $\Omega$ is the order of magnitude of the relaxing interaction and $\tau _c$ is generally taken as  the momentum relaxation time \cite{Dyakonov1971}.  Here, for a hopping transport, it has been pointed out that relaxation only occurs during the hopping process and that $\tau _c$ is the hopping time \cite{shklovskii2006}. In this case, it  seems clear that $\tau _c$ should decrease with increasing excitation power, because of the increase of the characteristic energy of the electrons in the fluctuations and possibly because of screening of the fluctuations by the photocarriers. This implies that the losses by spin relaxation are smaller at high  excitation power, at which  $\tau _c$ is relatively small. In the same way, this model explains that the polarization losses are smaller for hot electrons, for which $\tau _c$ is smaller than for Fermi edge electrons. Note finally that the presence of spin orientation up to 20 $\mu$m implies that photocarriers are transported over this distance. The mechanisms for this charge transport will be discussed in [II]. \

\subsection{Spatial dependence of the main emission line (M)}

\begin{figure}[tbp]
\includegraphics[clip,width=8.7 cm] {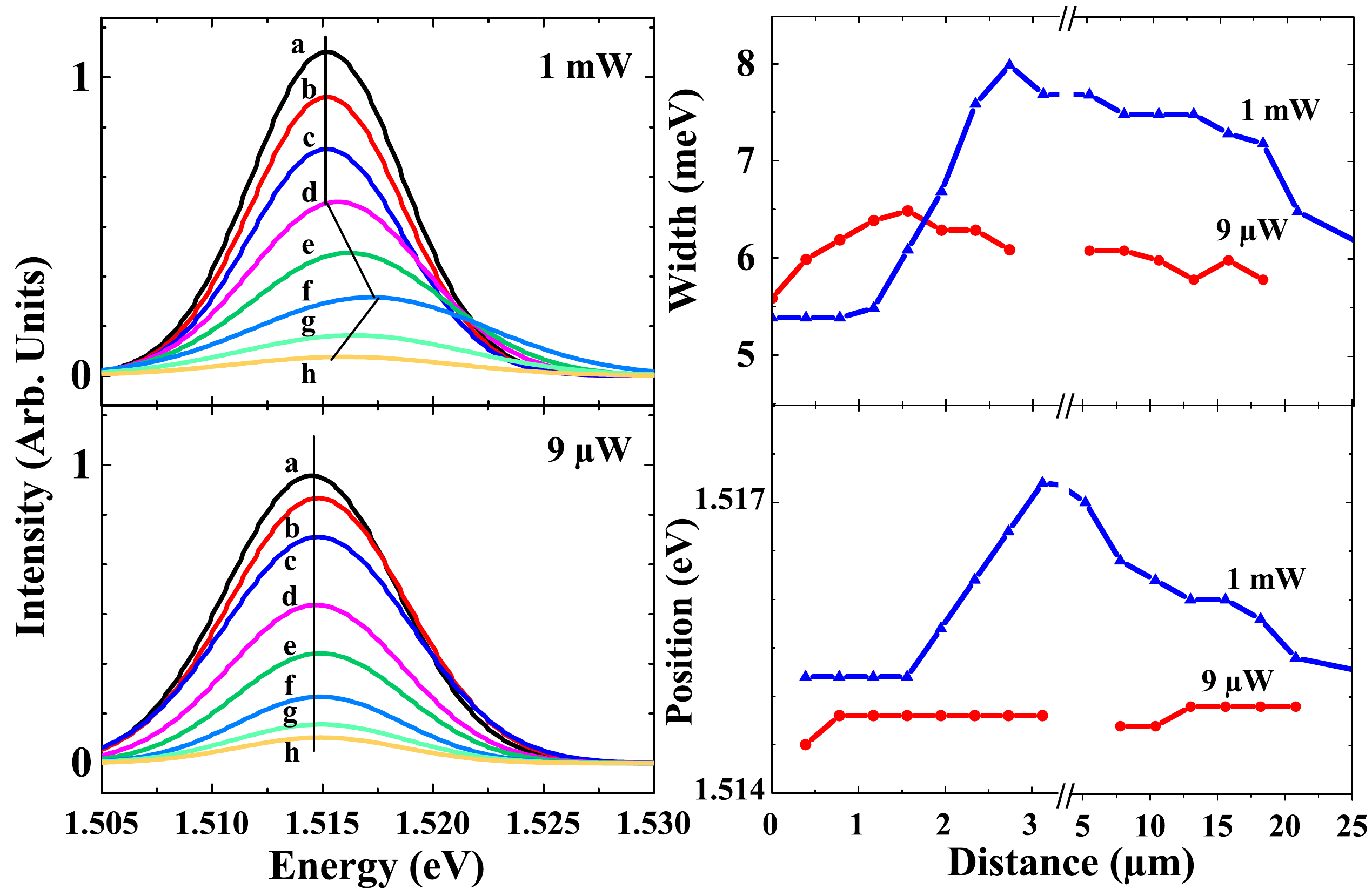}
\caption{The bottom left panel shows the shape of component M used to fit the intensity spectra for an excitation power of 9 $\mu$W at the excitation spot (Curve a) and at a distance from the excitation spot of 0.78 $\mu$m (Curve b),  1.56 $\mu$m (Curve c), 5.2 $\mu$m (Curve d), 10.4 $\mu$m (Curve e), 13 $\mu$m (Curve f), 15.6 $\mu$m (Curve g),  18.2 $\mu$m (Curve h). The top left panel shows the  same results for an excitation power of 1 mW at the excitation spot (Curve a) and at a distance from the excitation spot of 0.78 $\mu$m (Curve b), 1.17 $\mu$m (Curve c), 1.56 $\mu$m (Curve d), 2.6 $\mu$m (Curve e), 5.2 $\mu$m (Curve f), 10.4 $\mu$m (Curve g). Also shown in Curve e is the intensity profile of the laser spot. The top left panel shows, for the two excitation powers, the dependence of the gaussian linewidth on distance, while the bottom left panel shows the distance dependence of the energy of component M. }
\label{Fig8}
\end{figure}

In this section, we discuss the converse effect of the photoelectron system on the other reservoirs, such as the Fermi sea and the photohole reservoir. Since possible perturbations will affect the characteristics of the band-to-band recombination, one shows in Fig. \ref{Fig8} the spatial dependence of line M, used in the decomposition of the sum spectra, for the mimimum and maximum excitation powers, respectively. The distance dependences of the linewidth and position of line M are summarized in the top right panel and bottom right panel of  Fig. \ref{Fig8}, respectively.  At low excitation power, as expected because of the low photoelectron concentration, the peak position and the width are nearly constant. In contrast, for the maximum power, the line characteristics exhibit a significant spatial dependence. Up to  1.5 $\mu$m, the modifications of both linewidth and peak position are undetectable. For larger distances up to 3   $\mu$m, the linewidth increases and the peak shifts to higher energy. Further evolution, up to 25 $\mu$m, consists in a progressive return to the values at $z=0$.\ 

The change of  width and peak position of line M  is not correlated with  the  modification of the temperature $T_e$ of the Fermi sea, shown in Fig. \ref{Fig3}, since $T_e$ is maximum at $z=0$ and constantly decreases up to 25 $\mu$m. These effects cannot be attributed to a change in lattice temperature which would, at variance with observations, change the energy of other lines such as the acceptor-related lines. Finally, these changes cannot be correlated with changes of the Fermi level since, in the spatial range  up to  2.5   $\mu$m in which line S is observed, the width and position of this line are independent on space, implying that both the electronic quasi Fermi level and the hole level given by Eq. \eqref{holeferm} weakly depend on space.\

These considerations allow us to exclude a modification of the Fermi sea as an explanation of the results of Fig. \ref{Fig8}. We rather believe that the change is caused by the modification of the photohole occupation probability of the valence band tail.  Calculations using Eq. \eqref{IMind} and model parameters have indeed predicted, as observed here, that the line peak shifts to high energy upon increase of the photocarrier concentration \cite{levanyuk1981}. \

\section{Conclusion}

We have investigated the spatial dependence of the luminescence intensity and difference spectra as a function of distance to the excitation spot  in HVPE-grown, plasma-passivated GaAs NWs, at 6K and as a function of excitation power. These  NWs have a n-type doping level in the  low  $ 10^{17}$ cm$^{-3}$ range, implying that significant  tails are present both for the conduction and the valence band. The nearbandgap luminescence line is decomposed into three components : i) a relatively broad band at an energy slightly smaller than the bandgap, caused by recombination of photoholes with intrinsic, spin-unpolarized photoelectrons; ii) at nearbandgap energy, emission caused by recombination of the spin-polarized photoelectrons with the photoholes  iii)  emission caused by recombination of hot spin-polarized electrons with the same photoholes.\

From the compared evolution of the two types of lines as a function of distance to the excitation spot, it is possible to follow the establishment of equilibrium between the photoelectron, intrinsic electron and photohole reservoirs. It is found that, after a distance from the excitation spot of the order of  2 $\mu$m, photoelectron charges are in equilibrium with intrinsic electrons. Upon increase of the excitation power, the dynamics for establishment of equilibrium becomes slower, especially near the excitation spot. In this case, modification of the hole energy distribution in the valence bandtail induces a broadening and a shift to high energy of the main luminescence emission (line M). This perturbation is visible after a characteristic distance of 1.5   $\mu$m, up to about 3  $\mu$m  from the excitation spot and progressively returns to normal.\

Finally, it is found that the photoelectron spins are not affected by the above establishement of equilibrium and that photoelectrons and the Fermi sea remain distinct spin reservoirs, although their charges are in thermodynamic equilibrium. At large excitation power, the photoelectron spin polarization is preserved up to a  record distance of 20 $\mu$m. The decrease of excitation power leads to an increase of the polarization losses. These losses are attributed to hopping relaxation. Achievement of spin transport over this record length  implies that such NWs are good candidates for spintronics applications.\

\noindent

\acknowledgments 

{This work was supported by Région Auvergne Rhône-Alpes (Pack ambition recherche; Convention  17 011236 01- 61617,  CPERMMASYF and LabExIMobS3 (ANR-10-LABX-16-01). It was also funded by the program Investissements d 'avenir of the French ANR agency, by the French governement IDEX-SITE initiative 16-IDEX-0001 (CAP20-25), the European Commission (Auvergne FEDER Funds).}
 
\bibliographystyle{apsrev}


\begin{thebibliography}{57}
\bibitem[{not({\natexlab{a}})}]{note10}
	\bibinfo{note}{Since the effective diameter of the NW bulk, after removal of
		the surface depletion, is about 4 times the Bohr radius $a_0^*$, screening is
		rather three-dimensional. However, the TF model is only indicative since it
		assumes that screening only induces an energy shift of the electronic states.
		This approximation does not hold here since localized states can be
		suppressed or added by the mobile carriers. As a result, a numerical
		self-consistent approach is necessary in order to correctly describe the
		screening of the fluctuations by mobile carriers}.
\bibitem[{not({\natexlab{b}})}]{note8}
	\bibinfo{note}{Since the NW diameter is smaller than both the excitation spot
		and the light absorption depth, this excitation power is equivalent to a very
		small effective power density of $\approx 30 W/$cm$^2$.}
\bibitem[{not({\natexlab{c}})}]{note9}
	\bibinfo{note}{The downward shift of the conduction band minimum is probably
		comparable with the Fermi energy, so that the Fermi level should lie near the
		unperturbed conduction band minimum.}
\end{thebibliography}

\end{document}